\documentclass[sn-mathphys,iicol]{sn-jnl}
 
\usepackage{slashed}

\jyear{2023}%

\begin{document}

\title[""]{Non-forward radiative corrections to electron-carbon scattering}

\author[1,2]{\fnm{M.} \sur{Mihovilovi\v{c}}}
\author[3]{\fnm{P.} \sur{Achenbach}}
\author[2]{\fnm{M.} \sur{Bajec}}
\author[4]{\fnm{T.} \sur{Beranek}}
\author[1]{\fnm{J.} \sur{Beri\v{c}i\v{c}}}
\author[5,6]{\fnm{J.~C.} \sur{Bernauer}}
\author[4]{\fnm{R.} \sur{B\"{o}hm}}
\author[7]{\fnm{D.} \sur{Bosnar}}
\author[4]{\fnm{M.} \sur{Cardinali}}
\author[8]{\fnm{L.} \sur{Correa}}
\author[1]{\fnm{L.} \sur{Debenjak}}
\author[4]{\fnm{A.} \sur{Denig}}
\author[4]{\fnm{M.~O.} \sur{Distler}}
\author[4]{\fnm{A.} \sur{Esser}}
\author[4]{\fnm{M.~I.} \sur{Ferretti~Bondy}}
\author[8]{\fnm{H.} \sur{Fonvieille}}
\author[9]{\fnm{J.~M.} \sur{Friedrich}}
\author[7]{\fnm{I.} \sur{Fri\v{s}\v{c}i\'{c}}}
\author[10]{\fnm{K.} \sur{Griffioen}}
\author[4]{\fnm{M.} \sur{Hoek}}
\author[4]{\fnm{S.} \sur{Kegel}}
\author[4]{\fnm{H.} \sur{Merkel}}
\author[4]{\fnm{D.~G.} \sur{Middleton}}
\author[4]{\fnm{U.} \sur{M\"{u}ller}}
\author[4]{\fnm{L.} \sur{Nungesser}}
\author[4]{\fnm{J.} \sur{Pochodzalla}}
\author[4]{\fnm{M.} \sur{Rohrbeck}}
\author[4]{\fnm{S.}  \sur{S\'anchez~Majos}}
\author[4]{\fnm{B.~S.} \sur{Schlimme}}
\author[4]{\fnm{M.} \sur{Schoth}}
\author[4]{\fnm{F.} \sur{Schulz}}
\author[4]{\fnm{C.} \sur{Sfienti}}
\author[1,2]{\fnm{S.} \sur{\v{S}irca}}
\author[1]{\fnm{S.} \sur{\v{S}tajner}}
\author[4]{\fnm{Y.} \sur{St\"{o}ttinger}}
\author[4]{\fnm{M.} \sur{Thiel}}
\author[4]{\fnm{A.} \sur{Tyukin}}
\author[4]{\fnm{M.} \sur{Vanderhaeghen}}
\author[4]{\fnm{A.~B.} \sur{Weber}}
\author[4]{\fnm{M.}~\sur{Weinriefer}}

\affil[1]{\orgname{Jo\v{z}ef~Stefan~Institute}, \postcode{SI-1000}, \city{Ljubljana}, \country{Slovenia}}

\affil[2]{\orgdiv{Faculty of Mathematics and Physics}, \orgname{University of Ljubljana}, \postcode{SI-1000}, \city{Ljubljana}, \country{Slovenia}}

\affil[3]{\orgname{Thomas Jefferson National Accelerator Facility}, \city{Newport News}, \postcode{VA 23606}, \country{USA}}

\affil[4]{\orgdiv{Institut f\"{u}r Kernphysik}, \orgname{Johannes Gutenberg-Universit\"{a}t Mainz}, \postcode{DE-55128}, \city{Mainz}, \country{Germany}}

\affil[5]{\orgname{Stony Brook University}, \city{Stony Brook}, \postcode{NY 11794}, \country{USA}} 

\affil[6]{\orgname{Riken BNL Research Center}, \city{Upton}, \postcode{NY 11793},\country{USA}}

\affil[7]{\orgdiv{Department of Physics}, \orgname{University of Zagreb}, \postcode{HR-10002}, \city{Zagreb}, \country{Croatia}}

\affil[8]{\orgname{Universit\'{e} Clermont Auvergne}, \orgdiv{CNRS/IN2P3, LPC, BP~10448}, \postcode{F-63000} \city{Clermont-Ferrand}, \country{France}}

\affil[9]{\orgname{Technische Universit\"{a}t M\"{u}nchen}, \orgdiv{Physik Department}, \postcode{85748} \city{Garching}, \country{Germany}}

\affil[10]{\orgname{The College of William and Mary}, \city{Williamsburg}, \postcode{VA 23187}, \country{USA}}


\abstract{Radiative corrections to elastic scattering represent an important part of the interpretation of electron-induced  nuclear reactions at small energy transfers, where they make for a dominant part of background. Here we present and validate a new event generator for mimicking QED radiative processes in electron-carbon scattering that exactly calculates the coherent sum of the Bethe-Heitler amplitudes for the leading diagrams and can be reliably employed for a more robust determination of inelastic cross-sections.}



\maketitle

\section{Introduction}\label{sec1}
The studies of QED radiative corrections in electron-induced nuclear reactions have a long history, starting in 1949 with the first calculation performed by Schwinger~\cite{Schwinger1949}. Radiative corrections have been most extensively studied for the case of $e$--$p$ scattering~\cite{Tsai1961,Yennie:1961ad, MoTsai1969, Maximon1969, Maximon2000, Vanderhaeghen2000}. The success of calculations in the interpretation of the virtual Compton scattering experiments~\cite{A1:2015wgf, A1:2019mrv} as well as in the  determination of the proton charge form-factors \cite{Mihovilovic2017, Mihovilovic2019} has motivated us to extend the effort also to the electron-nucleus scattering. 

We are particularly interested in a precise understanding of the radiative corrections in electron-induced reactions on ${}^{12}\mathrm{C}$, which are relevant for all experiments that are aiming at measuring cross-sections with the systematic precision at the level of a few percent. One such example is the determination of the transition form-factors of the excited states of carbon, in particular the Hoyle state. The measured cross-sections for these states can be strongly contaminated by the radiative tail of the dominant elastic peak~\cite{Chernykh:2010zu} and could limit the precision of the extracted form-factors and thus the efforts to understand the structure of the Hoyle state.

A detailed description of the QED radiative corrections is also important to ensure a reliable analysis of scattering data that are relevant for studying neutrinos. The interpretation of the signals measured in neutrino experiments relies on Monte-Carlo (MC) simulations of neutrino interactions with nuclei inside the detector volume. Hence, a tremendous effort is currently being invested to develop theoretical models and corresponding MC event generators that are capable of providing a full description of the neutrino-nucleus cross-sections for energies  between a few hundred $\mathrm{MeV}$  and a few $\mathrm{GeV}$. One of the most recent nuclear models employed in the neutrino generators is the SuSAv2-MEC model~\cite{Megias:2016lke}, which has been trained on the available electron-carbon scattering data~\cite{RevModPhys.80.189}. However, the available spectra show that the quasi-elastic data collected at small electron scattering angles are contaminated by the radiative tail of the elastic peak as well as by the excited states of carbon, which in turn causes models to overestimate the cross-sections at these kinematic configurations. To overcome this problem, a reliable simulation of the radiative corrections going beyond the peaking approximation is needed.   

A precise understanding of the radiative corrections for elastic electron-carbon scattering also opens a possibility of using thin plastic (polyethylene) foils as effective hydrogen targets in the proton-charge form-factor measurements at low values of four-momentum transfers $(Q^2\approx 10^{-3}\,\mathrm{GeV}^2/c^2)$, where the cross-sections for nuclear excitations and quasi-elastic scattering are small. Hence, the experimental background of the hydrogen elastic cross-section measurement would be dominated by events from the reaction ${}^{12}\mathrm{C} + e \rightarrow {}^{12}\mathrm{C}+ e' + \gamma$.  If a simulation of such QED processes with a relative precision of  $1\,\mathrm{\%}$ would be available for background subtraction in regions where $E_{e'}^{\mathrm{elastic}} - E_{e'} \gg 1\,\mathrm{MeV}$, then the use of a solid-state target would represent a significant experimental advantage compared to the use of traditional extended cryogenic targets~\cite{Mihovilovic2019}.

\section{Radiative tail for elastic scattering}
\begin{figure}[ht]
\begin{center}
 \includegraphics[width=0.5\textwidth]{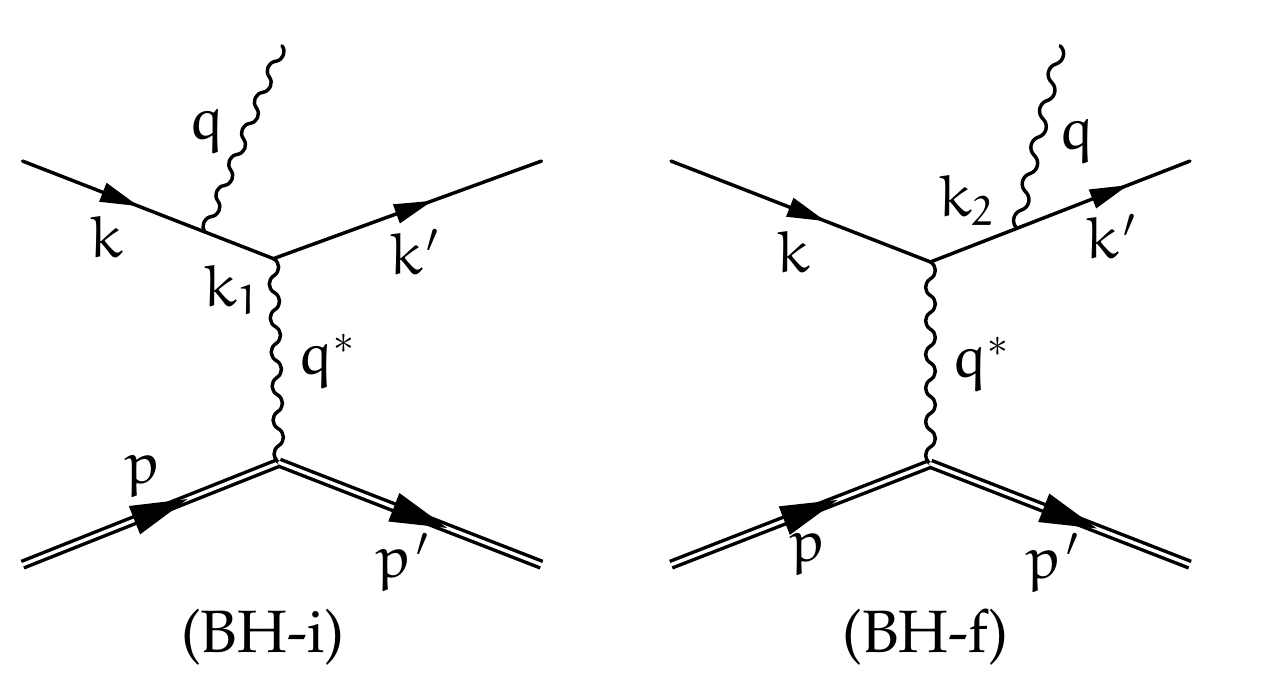}
 \caption{Feynman diagrams for Bethe-Heitler processes for scattering of an electron from a nucleus. In these reactions an electron emits a real photon before (BH-i) or after (BH-f) the interaction with the nucleus.
 \label{fig_BHDiagrams}}
 \end{center}
\end{figure}

The radiative tail for elastic scattering observed in the measured spectra arises from processes where one or more photons are radiated from the electron. The dominant contributions are represented by two Bethe-Heitler diagrams~\cite{Vanderhaeghen2000} shown in Fig.~\ref{fig_BHDiagrams}: the initial-state radiation diagram (BH-i) describes the process where the incident electron emits a real photon before interacting with the nucleus, while the final state radiation diagram (BH-f) corresponds to the reaction where the real photon is emitted after the interaction with the nucleus. 
Since in either case the emitted photon carries away a fraction of the energy,  the energy of the scattered electron, $E_{e'}$, is lower than the energy of the  elastically scattered electron, $E_{e'}^{\mathrm{elastic}}$. Consequently, some of the electrons miss the finite energy acceptance of the detector, causing the measured cross-section $({d\sigma}/{d\Omega})_\mathrm{exp.}$ to be smaller than the elastic cross-section $({d\sigma}/{d\Omega})_0$ (including next-order virtual corrections~\cite{Tsai1961, Bernauer2014}). This difference is encompassed in the radiative correction factor $\mathrm{e}^{\delta_R}$, such that:
\begin{eqnarray}
\left(\frac{d\sigma}{d\Omega}\right)_\mathrm{exp.}  =  \left(\frac{d\sigma}{d\Omega}\right)_\mathrm{0} \mathrm{e}^{\delta_R}\,.\label{eq:cs1}
\end{eqnarray}
The detailed calculations of the real-photon radiative corrections are presented in~\cite{Vanderhaeghen2000} and lead to the following expression for  $\delta_R$:
\begin{eqnarray}
  \delta_{R} &=& \delta_{R1} + \delta_{R2}\,,\nonumber\\
  \delta_{R1} &=& \frac{\alpha}{\pi} \ln\left( \frac{\eta^2\Delta E_{e'}^2}{E_e E_{e'}}\right)\left[ \ln\left(\frac{Q^2}{m_e^2}\right)-1\right]\,,\nonumber\\
  \delta_{R2} &=& \frac{\alpha}{\pi}\left\{ \frac{1}{2}\ln^2\frac{E_{e'}}{E_{e}} + \frac{1}{2}\ln^2\left(\frac{Q^2}{m_e^2}\right)- \right. \nonumber \\
  &&\left. \frac{\pi^2}{3} + \Phi\left(\cos^2\frac{\theta_{e'}}{2}\right) \right\}\,,  \label{eq_deltaR}
\end{eqnarray}   
where $\alpha$ is the fine-structure constant, $m_e$ is the mass of the electron, 
$\eta = {E_e}/E_{e'}^{\mathrm{elastic}}$ is the recoil factor, $\Delta E_{e'} = E_{e'}^{\mathrm{elastic}} - E_{e'}$ is the maximal energy loss of the scattered electron, and $\Phi(x)$ is the Spence function, defined by:
\begin{eqnarray}
\Phi(x) = -\int_0^x \frac{\ln\lvert 1-y\rvert}{y}dy\,.\nonumber
\end{eqnarray}
By introducing 
\begin{eqnarray}
	a = \frac{\alpha}{\pi}\left[\ln\left(\frac{Q^2}{m_e^2}\right)-1\right]\, \nonumber 
\end{eqnarray}
and $\Delta E_{e} = \eta^2\Delta E_{e'}$ as the energy loss of the incoming electron in the (BH-i) process~\cite{Vanderhaeghen2000}, the radiative correction factor (\ref{eq:cs1}) can be written as:
\begin{eqnarray}
\mathrm{e}^{\delta_R} &=& \eta^a \left(\frac{\Delta E_{e'}}{E_{e'}}\right)^{2a} \mathrm{e}^{\delta_{R2}}\nonumber\\
&=& \left(\frac{\Delta E_e}{E_e}\right)^a \left(\frac{\Delta E_{e'}}{E_{e'}}\right)^a \mathrm{e}^{\delta_{R2}}\,. \label{eq:cs2}
\end{eqnarray}
 For a more direct insight into the energy dependence of radiative corrections one needs to calculate the cross-section that depends also on $\Delta E_{e'}$. This is achieved by differentiating (\ref{eq:cs2}):
\begin{eqnarray}
\frac{d\sigma}{d\Omega_{e'}d\Delta E_{e'}} &=& \left(\frac{d\sigma}{d\Omega}\right)_\mathrm{0} \frac{2a \eta^a }{\Delta E_{e'}} \left(\frac{\Delta E_{e'}}{E_{e'}} \right)^{2a} \mathrm{e}^{\delta_{R2}}\label{eq:cs3}
\end{eqnarray}
From the obtained expression we can see that for the typical values of $a\approx 1\,\mathrm{\%}$, the term $(\Delta E_{e'}/E_{e'})^{2a}$  is approximately $1$, hence the radiative tail of the elastic peak has a $1/\Delta E_{e'}$ dependence.

\section{Peaking approximation}
\label{sec:PA}
The quantities $(\Delta E_e/E_e)^a$ and $(\Delta E_{e'}/E_{e'})^a$ represent the fractions of incoming and outgoing electrons which have lost an energy between 0 and $\Delta E_{e}$ or $\Delta E_{e'}$, respectively, and can be interpreted as cumulative functions of probability distributions $I$, such that:
$$
\int_{0}^{\Delta E_{x}} I(E_{x}, \Delta E_{x}, a) d \Delta E_{x} = (\Delta E_x/E_x)^a\,,
$$
where $x=e$ or $e'$. The distributions $I$ that satisfy this condition are:
\begin{eqnarray}
I(E_x, \Delta E_{x}, a) = \frac{a}{\Delta E_{x}} \left(\frac{\Delta E_{x}}{E_{x}} \right)^a\,. \label{eq:dist1}
\end{eqnarray}
Eqs.~(\ref{eq:cs3}) and (\ref{eq:dist1}) give us an idea how to mimic radiative corrections in a simulation~\cite{Vanderhaeghen2000}. First we generate the energy loss $\Delta E_{e}$ for an incoming electron with an initial energy $E_{e}$ according to the distribution $I(E_e, \Delta E_{e}, a)$. Then we reduce its energy to $\tilde{E}_{e} = E_{e} -\Delta E_{e}$ and use the corrected energy to calculate the energy of the elastically scattered electron: 
$$
\tilde{E}_{e'}^{\mathrm{elastic}} = \frac{\tilde{E}_{e}}{1+ \frac{\tilde{E}_{e}}{M_{{}^{12}\mathrm{C}}}(1-\cos{\theta_{e'}})}\,,
$$
where $M_{{}^{12}\mathrm{C}}$ is the mass of the carbon nucleus. In the third step we generate the energy loss of the scattered electron according to distribution $I(\tilde{E}_e, \Delta E_{e}, a)$ and calculate the final energy of scattered electron as $E_{e'} = \tilde{E}_{e'} -\Delta E_{e'}$. In order to generate events according to (\ref{eq:dist1}), the following prescription can be used:
\begin{eqnarray}
  \Delta E_{x}(r) = E_{x}r^{\frac{1}{a}}\,,\label{eq_generator1}
\end{eqnarray}
where $r$ is a random number uniformly distributed over $[0,1]$.
To calculate $a$ one needs $Q^2=4E_eE_{e'}\sin^2{(\theta_e/2)}$, which becomes known only after the energy of the scattered electron has been generated. However, since $a$ changes only slowly with $Q^2$, one typically considers a value for elastic electron scattering kinematics, $Q_{\mathrm{elastic}}^2 = 4E_eE_{e'}^{\mathrm{elastic}}\sin^2{(\theta_e/2)}$. 

On the other hand, for the calculation of the charge form-factor and cross-section $({d\sigma}/{d\Omega})_\mathrm{0}$ one should use the value of the four-momentum transfer at the vertex: $Q^2_{\mathrm{vertex}} = 4\tilde{E}_e\tilde{E}_{e'}^{\mathrm{elastic}}\sin^2{(\theta_e/2)}$. This way one considers also the changes in the cross-section due to the change of the energy of the incoming electron.

The devised numerical model assumes that the real photons are emitted in the direction of the incoming and scattered electrons, which is known as the (angular) peaking approximation. 

\section{Full simulation}
\label{sec:fullsim}
The peaking approximation has been traditionally used in the analysis of nuclear scattering experiments. However, it is known from the study of electron-proton scattering~\cite{Vanderhaeghen2000} that the peaking approximation starts to significantly overestimate the cross-section when $\Delta E_{e'} > 10\,\mathrm{MeV}$. This is not a problem when measuring elastic cross-sections but can represent an important source of systematic uncertainty when such a simulation is used to estimate the Bethe-Heitler background. For an adequate description far away from the elastic line it is crucial to consider cross-section contributions to the $e^8$-order. To achieve this goal, a more sophisticated Monte-Carlo simulation has been devised, which employs an event generator that exactly calculates the coherent sum of the amplitudes for the leading ($e^3$-order) diagrams shown in Fig.~\ref{fig_BHDiagrams} and then adds next-order contributions as effective corrections to the cross-section. 

\subsection{Reaction kinematics}
In the reaction ${}^{12}\mathrm{C} + e \rightarrow {}^{12}\mathrm{C} + e' + \gamma$ the initial electron with four-momentum $k$ scatters from the nucleus with the four-momentum $p$. In the process, a real photon with four-momentum $q$ is produced that leaves the reaction together with the scattered electron ($k'$) and the recoiled carbon nucleus ($p'$). For the reaction the following four-momentum conservation relation holds:
\begin{eqnarray}
 k - {k'} -q = {q^*} = {p'} - p\,, \nonumber 
\end{eqnarray}
Here  ${q^*}$ represents the four-momentum of the virtual photon exchanged between the lepton and the nucleus. We also define four-vectors $k_1$ and $k_2$ as the momenta of the internal electron legs in the considered Feynman diagrams:
\begin{eqnarray}
	k_1 = k - q\,,\qquad k_2 = {k'} + q \,.\nonumber 
\end{eqnarray} 

\subsection{Inelastic cross-section}
The exclusive cross section for the  reaction ${}^{12}\mathrm{C} + e \rightarrow  {}^{12}\mathrm{C}+ e' + \gamma$ in the laboratory frame (LAB) can be written as~\cite{Peskin:1995ev}:
\begin{eqnarray}
d\sigma &=& \frac{\overline{\lvert M_{fi} \rvert^2}}{4M_{{}^{12}\mathrm{C}} \mid\vec{k}\mid}  \frac{d^3\vec{k}'} 
{(2\pi)^3 2k_0'} \frac{d^3\vec{q}}{(2\pi)^3 2q_0} \frac{d^3 \vec{p}'}{(2\pi)^3 2 p_0' } \times \nonumber \\ \nonumber \\ 
&& (2\pi)^4 \delta^4(p+ k - {p'} - {k'} - q)\,, \nonumber
\end{eqnarray}
where $\overline{\lvert M_{fi} \rvert^2}$ represents the spin-averaged square of the reaction amplitude. Integrating the expression over the three-momentum of the undetected (recoiled) nucleus and considering $d^3\vec{k}' = \lvert\vec{k}'\rvert k_0' dk_0' d\Omega_{k'}$, one gets the six-fold differential cross-section:
\begin{eqnarray}
	\frac{d^6\sigma}{dk_0' d\Omega_{k'}  dq_0 d\Omega_{q}}  &&= \frac{\overline{\lvert M_{fi}\rvert^2}}{32 M_{{}^{12}\mathrm{C}}} \frac{1}{(2\pi)^5} \frac{\lvert\vec{k}'\rvert}{\lvert\vec{k}\rvert}\lvert\vec{q}\rvert \times \nonumber\\
   && \frac{\delta(p_0 + k_0 - p_0' - k_0' - q_0)}{p_0'}\,. \label{eq:cs4}
\end{eqnarray}
To obtain the inclusive three-fold differential cross-section ${d^3\sigma}/{dk_0' d\Omega_{k'}}$ that can be compared to the experiment, where only the electron is detected, (\ref{eq:cs4}) needs to be integrated over the energy and angles of the emitted real photon. Although the LAB coordinate system, where $k_0 = E_{e}$ and $k_0' = E_{e'}$, is the natural system of the experiment, it is more convenient to do this integration in the center-of-mass coordinate system (CMS). Therefore, the second part of the cross-section calculation including the amplitude is calculated in the CMS, which is defined by the real photon ${q}^\mu$ and recoiled nucleus $p'$ as:
\begin{eqnarray}
\vec{p'}_{\mathrm{CMS}} + \vec{q}_{\mathrm{CMS}}  = 
\vec{k}_{\mathrm{CMS}} - \vec{k'}_{\mathrm{CMS}} + \vec{p}_{\mathrm{CMS}} = 0\,. 
\nonumber 
\end{eqnarray} 
To perform the transformations from the LAB to  CMS,  a four-vector
$z \equiv k - k' + p$ is introduced, defining how 
the CMS coordinate system moves with respect to the LAB system. With $z$, any four-momentum vector $x^\mu$ can be transformed from LAB to CMS by using the Lorentz transformation
\begin{eqnarray}
x^\mu_\mathrm{CMS} = \Gamma^\mu_\nu x^\nu_{\mathrm{LAB}}\,, \qquad \Gamma^\mu_\nu = \left( 
\begin{array}{cc}
\gamma & -\gamma \vec{\beta}^T \\
-\gamma \vec{\beta} & 1 + \frac{\gamma-1}{\beta^2}\vec{\beta}\vec{\beta}^T
\end{array}
\right)\,, \nonumber
\end{eqnarray}
where $\vec\beta$ is the relative velocity of the CMS, 
$\vec\beta = \vec{z} / z^0$ and $\gamma = 1/\sqrt{1-\beta^2}$.
Now, considering that in the CMS the $\delta$-function can be written as:
\begin{eqnarray}
	\delta(p_0 + k_0 - p_0' - k_0' - q_0) = \underbrace{\frac{p_0'}{q_0 + p_0'}\delta(\lvert\vec{q}\rvert - \lvert\vec{p}'\rvert)}_{\mathrm{CMS}}\,,\nonumber
\end{eqnarray} 
the cross-section~(\ref{eq:cs4}) transforms into:
\begin{eqnarray}
&&\frac{d^3\sigma}{ dk_0' d\Omega_{k'} } = \frac{1}{32 (2\pi)^5 M_{{}^{12}\mathrm{C}}} \frac{\lvert\vec{k}'\rvert }{\lvert\vec{k}\rvert} \times \nonumber \\ \nonumber \\
&&\underbrace{\int \int   
\overline{\lvert M_{fi}\rvert^2}~\lvert\vec{q}\rvert\frac{p_0'}{q_0 + p_0'}\frac{\delta(\lvert\vec{q}\rvert - \lvert\vec{p}'\rvert)}{p_0'} d\Omega_{q} dq_0}_{\mathrm{Calculated~in~CMS}}\,. \nonumber
\end{eqnarray}
Integrating this formula over the energy of the emitted photon and introducing the Mandelstam invariant $s = (k-k' + p)^2 = (p' + q)^2 = z^2$, one obtains the following expression for the cross section:
\begin{eqnarray}
\frac{d^3\sigma}{ dk_0' d\Omega_{k'}}  &=& \frac{1}{64 (2\pi)^5 M_{{}^{12}\mathrm{C}}}\frac{\lvert\vec{k}'\rvert}{\lvert\vec{k}\rvert} \times \nonumber\\
&&\underbrace{\int 
\overline{\lvert M_{fi}\rvert^2}\frac{s - M_{{}^{12}\mathrm{C}}^2}{s}
d\Omega_{q}}_{\mathrm{Calculated~in~CMS}} \,.\nonumber
\end{eqnarray}
The following relations, valid in the CMS, were considered:
\begin{eqnarray}
	s &=&  \left(p_0' + q_0' \right)^2 + (\vec{p}' + \vec{q}')^2 = \left(p_0' + q_0' \right)^2\,,\nonumber\\
	s &=& M_{{}^{12}\mathrm{C}}^2 + 2p_0'q_0 - 2\vec{p}'\cdot\vec{q} = M_{{}^{12}\mathrm{C}}^2 + 2q_0(q_0 + p_0)\,,\nonumber\\
 	&&\frac{\lvert\vec{q}\rvert}{q_0+ p_0'} = \frac{q_0(q_0 + p_0')}{(q_0 + p_0')^2} = \frac{s-M_{{}^{12}\mathrm{C}}^2}{2s}\,.\nonumber
\end{eqnarray}
The integration over the solid angle is performed by means of the Monte-Carlo simulation itself. However, the simulation typically generates photon angles in the LAB system, while integration requires angles in the CMS. Hence, a Jacobian is needed that relates $(\Delta\Omega_q)_{\mathrm{LAB}}$ to $(\Delta\Omega_q)_{\mathrm{CMS}}$:
\begin{eqnarray}
(\Delta\Omega_q)_{\mathrm{CMS}} = \frac{(\partial\Omega_q)_{\mathrm{CMS}}}{(\partial\Omega_q)_{\mathrm{LAB}}}
(\Delta\Omega_q)_{\mathrm{LAB}}
\end{eqnarray}
This transformation can be implemented by using the following approximation. First, we define vectors $\vec{a}_\mathrm{LAB}$ and $\vec{b}_\mathrm{LAB}$ of length $\epsilon \approx 10^{-7}$ in LAB that are perpendicular to $\vec{q}$ and perpendicular to each other. Then we transform these two vectors to CMS, 
obtaining $\vec{a}_\mathrm{CMS}$ and $\vec{b}_\mathrm{CMS}$. The ratio between the areas determined by these new vectors in both coordinate systems gives us the estimate for the Jacobian:
\begin{eqnarray}
 \frac{(\partial\Omega_q)_{\mathrm{CMS}}}{(\partial\Omega_q)_{\mathrm{LAB}}} \approx 
\frac{dS_{\mathrm{CMS}}}{dS_{\mathrm{LAB}}} =
 \frac{\lvert\vec{a}_\mathrm{CMS}\times \vec{b}_\mathrm{CMS}\rvert}{\lvert\vec{a}_\mathrm{LAB}\times \vec{b}_\mathrm{LAB}\rvert}\,. \nonumber
\end{eqnarray}
The inclusive cross section for ${}^{12}\mathrm{C} + e \rightarrow  {}^{12}\mathrm{C}+ e' + \gamma$, which depends only on the three-momentum of the scattered electron, and is integrated over $d\Omega_q$ in the LAB coordinate system, then equates to:
\begin{eqnarray}
\frac{d^3\sigma}{ dk_0' d\Omega_{k'} } &=& \int d\Omega_{q} \frac{\overline{\lvert M_{fi}\rvert^2}}{64 (2\pi)^5 M_{{}^{12}\mathrm{C}}}\frac{\lvert\vec{k}'\rvert}{\lvert\vec{k}\rvert}
 \left( \frac{s - M_{{}^{12}\mathrm{C}}^2}{s}\right) \times \nonumber \\
 &&\frac{\lvert\vec{a}_\mathrm{CMS}\times \vec{b}_\mathrm{CMS}\rvert}{\lvert\vec{a}_\mathrm{LAB}\times \vec{b}_\mathrm{LAB}\rvert} \,.
 {}\label{eq:cs5}
\end{eqnarray}


\subsection{Amplitude for ${}^{12}\mathrm{C} + e \rightarrow  {}^{12}\mathrm{C}+ e' + \gamma$}
The amplitudes for the Bethe-Heitler diagrams shown in Fig.~\ref{fig_BHDiagrams}
can be written as:
\begin{eqnarray}
&&M_{\mathrm{BH-i}} = \left[i Z_{{}^{12}\mathrm{C}}\,e_0 (p+p')^{\kappa} F({q^{*}}^2)\right]\left(\frac{-ig_{\mu\kappa}}{{q^*}^2}\right) \times  \nonumber \\ 
&&\left[\overline{u}_{s'}(k'){\epsilon^*}_\nu \left(-ie_0\gamma^\mu\right)\frac{i(\slashed{k_1}+m_e)}{{k_1}^2-m_e^2}  \left(-ie_0\gamma^\nu\right)u_s(k) \right]
 \,,\nonumber\\\nonumber\\
&&M_{\mathrm{BH-f}} = \left[i Z_{{}^{12}\mathrm{C}}\,e_0 (p+p')^{\kappa}F({q^{*}}^2)\right]\left(\frac{-ig_{\mu\kappa}}{{q^*}^2}\right)\times \nonumber \\
&&\left[\overline{u}_{s'}(k'){\epsilon^*}_\nu \left(-ie_0\gamma^\nu\right)\frac{i(\slashed{k_2}+m_e)}{{k_2}^2-m_e^2}
\left(-ie_0\gamma^\mu\right)u_s(k) \right] \,, \nonumber
\end{eqnarray}
where we were using abbreviated notation $\slashed{k} = k_\alpha\,\gamma^\alpha$ and $Z_{{}^{12}\mathrm{C}} = 6$ is the charge number of ${}^{12}\mathrm{C}$. The terms in the square brackets represent the electron and carbon currents together with the  electron and nucleus vertex factors, respectively. We considered that the electron is a spin-$1/2$ particle, while the carbon nucleus is spinless.  Here $e_0$ is the elementary charge, while $F({q^{*}}^2)$ represents the nuclear charge form-factor.  The terms $\frac{i(\slashed{k_i}+m_e)}{{k_i}^2-m_e^2}$ correspond to internal electron propagators with momentum $k_i$, while $\frac{-ig_{\nu\kappa}}{{q*}^2}$ is the virtual photon propagator. 
The Dirac bispinors $u$ and $\overline{u}$ correspond to incoming and outgoing fermions with spin  ($s,s'=\uparrow,\downarrow$). The two particle states can be written as: 
\begin{eqnarray}
u_{\uparrow}(k) = \sqrt{k_0+m_ec^2} \left(
\begin{array}{c}
1 \\
0 \\
\frac{c k_z}{k_0+m_ec^2} \\ 
\frac{c(k_x + ik_y)}{k_0+m_ec^2}
\end{array}
\right)\,, \nonumber\\
u_{\downarrow}(k)  = \sqrt{k_0+m_ec^2} \left(
\begin{array}{c}
0 \\ 
1 \\ 
\frac{c(k_x - ik_y)}{k_0+m_ec^2}\\  
-\frac{c k_z}{k_0+m_ec^2}
\end{array}
\right)\,,\nonumber
\end{eqnarray}  
with the normalization $\overline{u}_s u_{s'} = 2k_0\delta_{s,s'}$. 

It needs to be stressed that in general the calculation of the one-photon emission amplitude should involve also Feynman diagrams where the carbon nucleus itself emits a real photon before or after the interaction with the electron. However, with respect to the Bethe-Heitler processes, these diagrams are suppressed approximately as $k^2/M_{{}^{12}\mathrm{C}}^2 \approx 10^{-4}$ and can thus be neglected.   

\subsection{Carbon charge form-factor}
\begin{figure}[ht]
\begin{center}
 \includegraphics[width=0.45\textwidth]{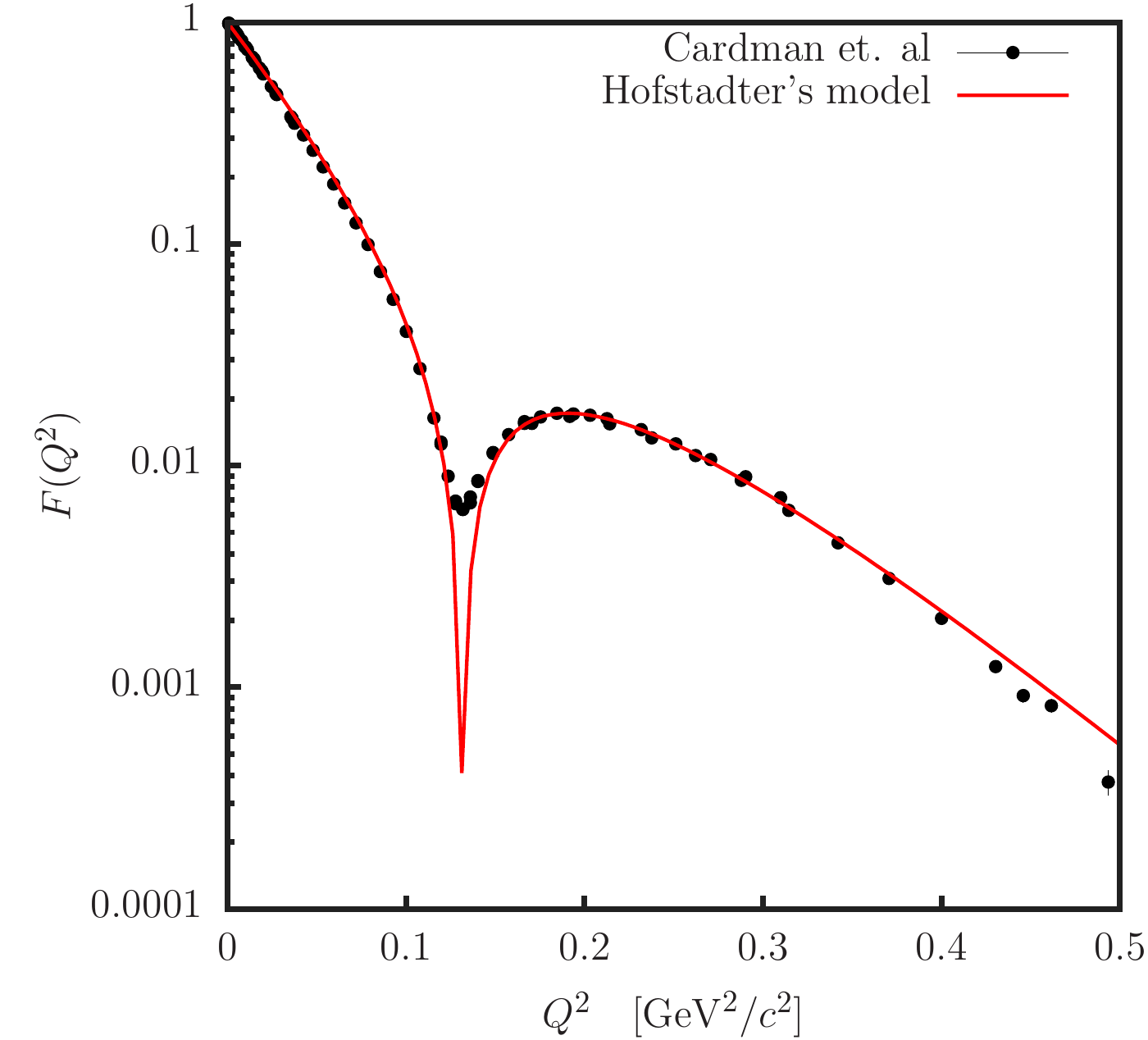}
 \caption{Hofstadter's parameterisation~\cite{Hofstadter:1956qs} of the carbon charge form-factor, $F(Q^2)$ superimposed to the existing measurements of Cardman~et~al.~\cite{Cardman:1980dja}.
 \label{fig_ElasticCS}}
 \end{center}
\end{figure}

The elastic interaction of electrons with spinless carbon nuclei through the exchange of virtual photons is governed by the charge form-factor $F(Q^2)$, which carries information about the charge distribution inside the nucleus. Various parameterizations of its charge distribution and form factor are available for carbon. In this work we considered the parameterization of Hofstadter~\cite{Hofstadter:1956qs}, 
$$
F(Q^2) = \left[1- \frac{\beta a^2Q^2 }{2k^2(2+3\beta)} \right]\exp{\left(-\frac{a^2Q^2}{4k^2}\right)}\,,
$$
where $k = \sqrt{3(2+5\beta)/2(2+3\beta)}$, $\beta = 4/3$ and $a = 2.4\,\mathrm{fm}$ are the parameters of the model~\cite{Hofstadter:1956qs}. With the exception of the region around the diffraction minimum at $Q^2 = 0.13\,\mathrm{GeV^2}/c^2$ the model describes the data \cite{Cardman:1980dja} reasonably well, with the relative precision better than $10\,\mathrm{\%}$ at $Q^2<0.35\,\mathrm{GeV^2}/c^2$. In the region of interest, $Q^2<0.003\,\mathrm{GeV^2}/c^2$, the agreement between the data and model is at the level of $0.5\,\mathrm{\%}$, see Fig.~\ref{fig_ElasticCS}.

\subsection{ The spin-averaged square amplitude}
The calculation of the spin-averaged square amplitude, $\overline{\lvert M_{fi} \rvert^2}$, is performed in several steps. In the first step a summation over the index $\mu$ is performed. For given spin states of electrons, the following sum is calculated:
\begin{eqnarray}
	M_{fi}^\nu &=& \frac{-ie_0^3}{{q^*}^2} \sum_\mu \overline{u}_{s'}(k') \mathrm{\Xi_e}^{\nu\mu}u_s(k)\>  g_{\mu\mu}\>
(p+p')_\mu\,,\nonumber 
\end{eqnarray}
where we have considered that $(p+p')^{\kappa}g_{\kappa\mu} = (p+p')_{\mu}$ and we have written the leptonic transition tensor as:
\begin{eqnarray}
	 \mathrm{\Xi_e}^{\nu\mu} &=& \gamma^\mu\left(\frac{i(\slashed{k_1}+m_e)}{{k_1}^2-m_e^2}\right) \gamma^\nu +  
\gamma^\nu \left(\frac{i(\slashed{k_2}+m_e)}{{k_2}^2-m_e^2}\right) \gamma^\mu\,.\nonumber
\end{eqnarray}
In the next step, the square of the amplitude is calculated and summation over indices $\nu$ and $\nu'$ is performed. Considering also polarization vectors of the emitted photon and summing over all its possible polarization states  $\lambda$, the following relation holds:
\begin{eqnarray}
	\lvert M_{fi}\rvert^2 &=& 
\sum_{\nu}\sum_{\nu'}\sum_\lambda{\epsilon_\nu^{(\lambda)}}^* {\epsilon_{\nu'}^{(\lambda)}} 
M_{fi}^{\nu}{M_{fi}^{\nu'}}^* \nonumber\\
&=& -\sum_\nu g_{\nu\nu} M_{fi}^{\nu}{M_{fi}^{\nu}}^*\,, \label{eq:ampl1}
\end{eqnarray}
where we considered the completeness relation  
$\sum_\lambda{\epsilon_\nu^{(\lambda)}}^* {\epsilon_{\nu'}^{(\lambda)}} = -g_{\nu \nu'}$. In order to obtain the final value of the spin-averaged square of the amplitude, Eq.~(\ref{eq:ampl1}) must be averaged over possible spin states $s$ of the initial electron and summed over possible spin 
states $s'$ of the scattered electron:
\begin{eqnarray}
\overline{\lvert M_{fi} \rvert^2} = \overline{\sum_{s=\uparrow,\downarrow}}
\sum_{s'=\uparrow,\downarrow} 
\lvert M_{fi}\rvert^2 \nonumber
\end{eqnarray}
The second summation over the spin states of the final lepton is done explicitly. Averaging over the spin states of the initial electrons, on the other hand, is achieved by randomly generating spins of the initial electrons for each process within the Monte-Carlo simulation.   

\subsection{Generating random events}
The derived formalism can now be used to construct the event generator.
In order for the simulation to work, we need to randomly generate four-vectors $k$, $k'$ and $q$ (in the LAB frame) for each simulated event. For that purpose we first select the direction and energy of the incident electrons. Traditionally the beam of electrons is pointing along the $z$-axis of the LAB coordinate system, hence, $k=(E_e,0,0,\sqrt{E_e^2-m_e^2})$.
In the next step, spherical angles $\theta_{e'}$ and $\phi_{e'}$ of the scattered electron are randomly generated. They are uniformly distributed over the (spherical) angular acceptance of the spectrometer (detector). Once the direction of the scattered electron is known, its maximal (elastic) energy can be determined by using energy and momentum conservation:
\begin{eqnarray}
	E_{e'}^{\mathrm{elastic}} = \frac{(-E_{e} - M_{{}^{12}\mathrm{C}})(E_{e}M_{{}^{12}\mathrm{C}} + m_{e}^2)}{(E_{e}^2 - m_e^2)\cos^2\theta_{e'} - (E_{e}+M_{{}^{12}\mathrm{C}})^2} - \nonumber\\
    - \frac{\sqrt{(E_{e}^2 - m_e^2)^2\cos^2\theta_{e'}
\left[M_{{}^{12}\mathrm{C}}^2 - \frac{1}{2}m_e^2(1-\cos 2\theta)\right]}}{(E_{e}^2 - m_e^2)\cos^2\theta_{e'} - (E_{e}+M_{{}^{12}\mathrm{C}})^2}\,.
\nonumber
\end{eqnarray}
In the next step we generate the energy loss of the scattered electron $\Delta E_{e'} \in [0, E_{e'}^{\mathrm{elastic}}]$ and calculate the energy of the detected electron $E_{e'} = E_{e'}^{\mathrm{elastic}} - \Delta E_{e'}$. The corresponding four-momentum vector of the scattered electron is  then:
\begin{eqnarray}
k' &=& (E_{e'},\vec{k})\,, \nonumber \\
\vec{k'} &=& \lvert \vec{k'}\rvert (\sin\theta_{e'}\cos\phi_{e'}, \sin\theta_{e'}\sin\phi_{e'},\cos\theta_{e'})\,, \nonumber \\
\lvert \vec{k'}\rvert &=&  \sqrt{E_{e'}^2 - m_{e}^2}\,. \nonumber
\end{eqnarray}
Finally we generate also the emitted real photon. First we determine its direction $q^\mu / \lvert q^\mu\rvert$ by randomly choosing its spherical angles $\cos\theta_q \in [-1,1]$ and $\phi_q\in [0,2\pi)$. Then, in the LAB system, where the initial carbon nucleus is at rest, $p =(M_{{}^{12}\mathrm{C}},\vec{0})$, the magnitude of the momentum of the real photon $\lvert\vec{q}\rvert$ can be determined by using 
\begin{eqnarray}
 \lvert\vec{q}\rvert= \frac{{\vec{q^*}}^2 - q^*_0 - 2q^*_0 M_{{}^{12}\mathrm{C}}}{2\left( -q^*_0 - M_{{}^{12}\mathrm{C}} + \vec{q^*} \cdot\frac{ \vec{q}}{ \mid\vec{q}\mid}\right)}\,. \nonumber
\end{eqnarray} 

\subsection{Priority sampling}
The generated energy-losses of the electron $\Delta E_{e'}$ should in principle be uniformly distributed on $[0, E_{e'}^{\mathrm{elastic}}]$. Similarly, the directions of the emitted photon, given by $(\theta_q,\phi_q)$, should uniformly cover the $4\pi$ solid angle. In this case the weight of each event in the simulated histograms is determined by the corresponding cross-section. However, such a procedure is very inefficient and would require huge samples of events in order to generate presentable distributions of physically interesting quantities, for instance, the distribution of scattering angle $\theta_{e'}$ at a chosen value of $E_{e'}$. 

From (\ref{eq:cs3}) we know that the energy loss of the electron behaves roughly as $\eta^a/\Delta E_{e'}\cdot(\Delta E_{e'}/E_{e'})^{2a}$. Hence, we can use prescription (\ref{eq_generator1}) to generate energy losses more efficiently and then divide out the approximate distribution (\ref{eq:cs3}) when calculating the weight (cross-section) of each generated event.

Similarly, we know that the probability for emission of a real photon depends strongly on its direction. The most probable directions are along the incident and scattered electron. The probability for emission of a photon then drops rapidly when moving away from  these two principal directions, which means that  a random generator that distributes photons uniformly over $4\pi$  would make the simulation very inefficient. Therefore, we rather generate photon angles according to a distribution that mimics the Bethe-Heitler cross-section. Assuming that the photon is emitted from the incoming and the outgoing electron with equal probabilities, the approximate 
Bethe-Heitler cross section (ABH) can be written as~\cite{Bernauer2014}:
\begin{eqnarray}
 \left(\frac{d\sigma}{d\Omega}\right)_{\mathrm{ABH}} = \frac{1}{2} \left(\frac{d\sigma}{d\Omega}\right)_{e} + 
\frac{1}{2} \left(\frac{d\sigma}{d\Omega}\right)_{e'}\,, \nonumber
\end{eqnarray}
where contributions from both the incident and the scattered electron can be expressed as:
\begin{eqnarray}
\left(\frac{d\sigma}{d\Omega}\right)_{x} &=& \frac{1}{N(E_x, \vec{x})} 
\frac{1-\cos^2\theta_{x\gamma}}{\left(\frac{E_{x}}{\lvert\vec{x}\rvert} - \cos \theta_{x\gamma} \right)^2}\,,\nonumber\\
N(E_x, \vec{x}) &=& -4 -2\frac{E_x}{\lvert\vec{x}\rvert}\ln\left(\frac{\frac{E_x}{\lvert\vec{x}\rvert} -1}{\frac{E_x}{\lvert\vec{x}\rvert} +1} \right)\,.\label{eq:approxBH} 
\end{eqnarray}
Here $\vec{e}=\vec{k}$, $\vec{e'}=\vec{k'}$ and $\theta_{x\gamma}$ is a polar angle between the emitted photon and one of the electrons.  A number generator for $\theta_{x\gamma}$ distributed according to (\ref{eq:approxBH}), can be written as 
\begin{eqnarray}
	\theta_{x\gamma} = F^{-1}(r)\,,\nonumber
\end{eqnarray}
where $r$ is random number uniformly distributed on $[0,1]$, and $F^{-1}(r)$ is the inverse of the cumulative function $F(\theta_{x\gamma})$ of the distribution (\ref{eq:approxBH}), which can be calculated as:
\begin{eqnarray}
F(\theta_{x\gamma}) &=& \int_{1}^{\cos{\theta_{x\gamma}}}\left(\frac{d\sigma}{d\Omega}\right)_x d\cos\theta = \nonumber\\
&=&\frac{1}{N(E_x, \vec{x})}\left[ \frac{1 - \left(\frac{E_x}{\lvert\vec{x}\rvert}\right)^2}{\frac{E_x}{\lvert\vec{x}\rvert} - \cos\theta_{x\gamma}} - \cos \theta_{x\gamma} - \right. \nonumber \\ 
&& \left . 2 \frac{E_x}{\lvert\vec{x}\rvert} \ln\frac{\frac{E_x}{\lvert\vec{x}\rvert}-\cos\theta_{x\gamma}}{\frac{E_x}{\lvert\vec{x}\rvert}+1}-2 + \frac{E_x}{\lvert\vec{x}\rvert}
\right]\,.\nonumber
\end{eqnarray}
The approximate Bethe-Heitler distribution  (\ref{eq:approxBH}) must of course 
be divided out before the calculation of the final cross-section, by multiplying the weight of each generated event with $\left(d\sigma/d\Omega\right)_{\mathrm{ABH}}^{-1} $. The cross-section (\ref{eq:approxBH})  does not depend on the azimuthal angle $\phi_{x\gamma} \in [0, 2\pi]$, hence, $\phi_{x\gamma}$ can be generated by using the standard uniform random generator:
\begin{eqnarray}
\phi_{x\gamma} = 2\pi r\,, \qquad r\in [0,1]\,.
\end{eqnarray} 

\subsection{Higher order corrections}
In order for the event generator to realistically mimic the radiative tail, 
higher-order corrections are considered next to the leading order Bethe-Heitler processes. They include two-photon emission diagrams as well as virtual vertex and vacuum polarisation corrections. These additional diagrams have been considered in the approximation of the elastic scattering as multiplicative factor to the cross-section.  The diagrams corresponding to the emission of two real soft photons have been considered via Eqs.~(\ref{eq:cs1}) and (\ref{eq_deltaR}). The virtual corrections have been considered through the multiplicative factor $\mathrm{e}^{\delta_{\mathrm{vert}}}/{(1-\delta_\mathrm{vac})^2}$ \cite{Vanderhaeghen2000}, where
\begin{eqnarray}
\delta_\mathrm{vert} &=& \frac{\alpha}{\pi}\left\{ \frac{3}{2}\ln\left(\frac{Q^2}{m^2}\right)-2-\frac{1}{2}\ln^2{\left(\frac{Q^2}{m^2}\right)} + \frac{\pi^2}{6}\right\}\,, \nonumber\\
\delta_\mathrm{vac} &=& \frac{\alpha}{\pi}\frac{2}{3}\left\{-\frac{5}{3} + \ln{\left(\frac{Q^2}{m^2}\right)}\right\}\,.  \nonumber   
\end{eqnarray}

\section{Experiment}
The measurement of the radiative tail has been performed at the Mainz Microtron (MAMI) in 2013 during the ISR experiment~\cite{Mihovilovic2017} using the spectrometer setup of the A1-Collaboration~\cite{Blomqvist}. In the experiment a point-like electron beam with energy of $E_{e} = 195\,\mathrm{MeV}$ was used in combination with a solid carbon target with surface density of $92\,\mathrm{mg/cm^2}$. For the cross-section measurements the single-dipole magnetic spectrometer~B was employed. It was positioned at a fixed angle of $\theta_{e'}=15.21^\circ$, while its momentum settings were adjusted to scan the complete radiative tail on the energy range between $E_{e'} = 40\,\mathrm{MeV}$ and $195\,\mathrm{MeV}$. The central momentum of each of the $19$ settings was measured with an NMR probe to a relative accuracy of $8\times10^{-5}$. The spectrometer is equipped with a detector package consisting of two layers of vertical drift chambers (VDCs) for tracking, two layers of scintillation detectors for triggering, and a threshold Cherenkov detector for identification of electrons and cosmic background suppression. 

The beam current was between $10\,\mathrm{nA}$ and $1\,\mu\mathrm{A}$ and was limited by the maximum rate allowed  in the VDCs ($\approx 1\,\mathrm{kHz/wire}$), resulting in raw rates up to $20\,\mathrm{kHz}$. The current was determined by a non-invasive fluxgate-magnetometer and from the collected charge of the stopped beam.

\section{Data Analysis}
The measured cross-sections were used to validate the new Monte-Carlo generator of radiative correction for electron-carbon scattering. The tests were performed at energies of scattered electrons $E_{e'}$ that are up to $150\,\mathrm{MeV}$ away from the elastic configuration.    

Before comparing the data to the simulation, the measured spectra had to be corrected for the inefficiencies of the detection system. The efficiencies of the scintillation detectors and the Cherenkov detector were determined to be $(99.8\pm0.2)\,\mathrm{\%}$ and  $(99.74\pm0.02)\,\mathrm{\%}$, respectively, and were considered as multiplicative correction factors to the measured distributions. The quality of the agreement between the data and simulation depends also on the momentum and spatial resolutions of the spectrometer. The relative momentum plus angular and vertex resolutions (FWHM) were determined to be $1.7\times 10^{-4}$, $3\,\mathrm{msr}$, and $1.6\,\mathrm{mm}$,
respectively.

A series of cuts were applied to the data in order to minimize the background.
First, a cut on the Cherenkov signal was applied to identify electrons,
followed by a cut on the nominal momentum and angular acceptance of the spectrometer to remove distortions caused by the inefficiencies at the edges of the detectors. At the end a  $\pm 10\,\mathrm{mm}$ cut on the vertex 
position was applied to remove contributions of the events that entered the spectrometer's acceptance by rescattering from the metal entrance flange of the spectrometer. Since these events have wrong scattering angles, they (virtually) appear in the spectra at large vertex positions and can be efficiently removed with the applied cut on the vertex position.  

Additionally, the external radiative corrections in the target material were considered using the formalism of Mo and Tsai~\cite{MoTsai1969}, while the collisional corrections were approximated by the Landau distribution~\cite{landau1944}. The uncertainty of the applied energy corrections was estimated to be smaller than $1\,\%$~\cite{MoTsai1969}.  
  
The cleaned event samples for each kinematic setting were corrected for the dead-time and prescale factors, weighted by the luminosity (determined from the beam current and target density) and then finally merged together to form a single spectrum for the comparison with the simulation.  

\begin{figure}[ht!]
 \includegraphics[width=0.48\textwidth]{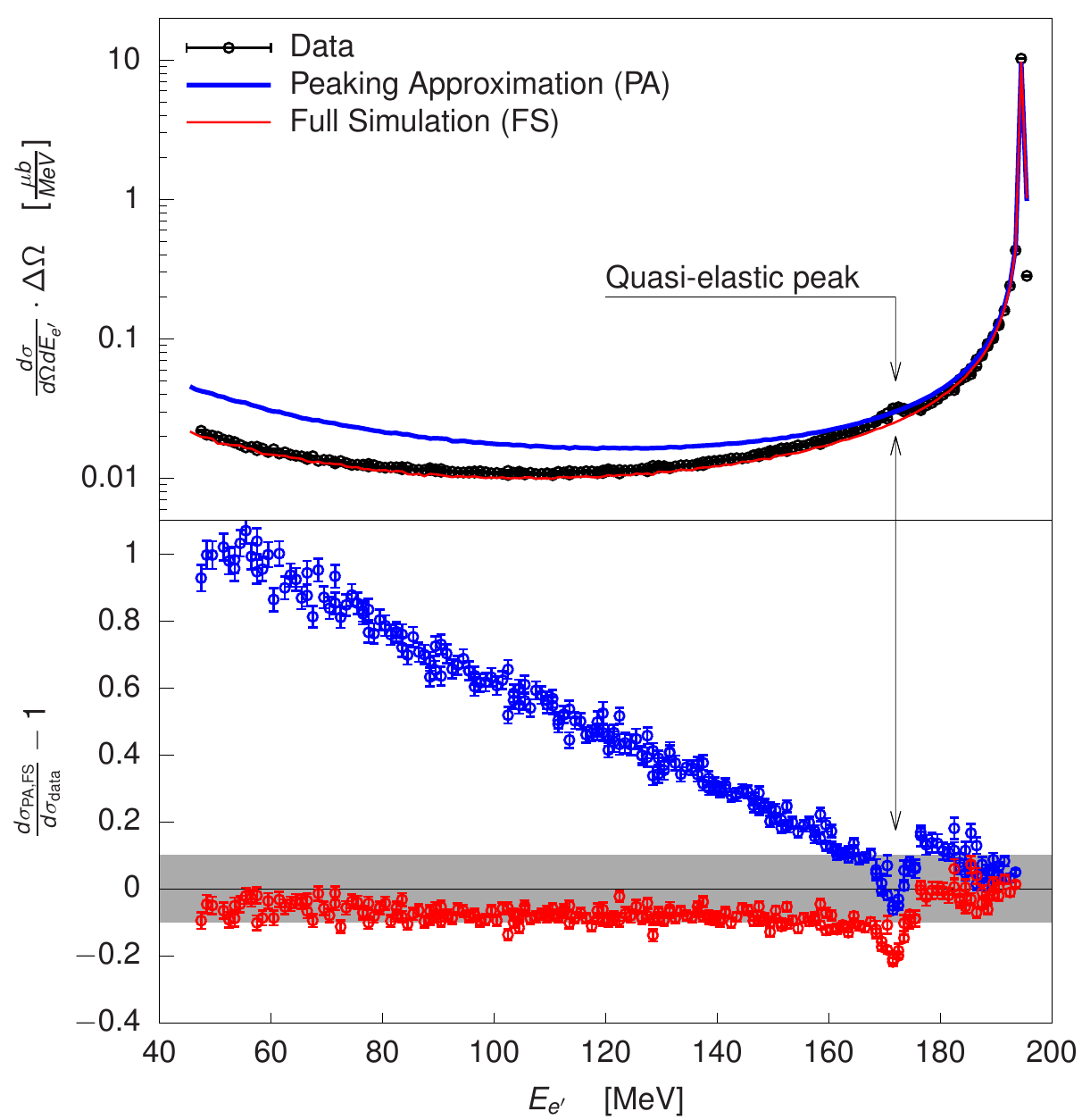}
 \caption{ 
Comparison of the data to the simulations. 
 {\bf Top:} The measured cross-section for inelastic scattering of electrons off carbon (black circles) for beam energy $195\,\mathrm{MeV}$ and scattering angle $15.2\,\mathrm{{}^\circ}$ together with the corresponding peaking approximation (blue line) and the detailed simulation discussed in this work (red line). A small peak at $172\,\mathrm{MeV}$ corresponds to the contribution of the quasi-elastic scattering process ${}^{12}\mathrm{C}+e\rightarrow e'+ p + {}^{11}\mathrm{B}$ to the measured spectrum. {\bf Bottom:} Relative differences between 
 the data and simulations at various energies of scattered electron $E_{e'}$. The blue points represent the comparison with the peaking approximation, while the red points correspond to the results for the newly developed detailed simulation. The gray band denotes the region where relative difference between the data and the simulation is smaller than $\pm 10\,\mathrm{\%}$.  Only statistical uncertainties are shown. The total systematic uncertainty is estimated to $1.2\,\mathrm{\%}$.
 \label{fig_ResultsEnergy}}
\end{figure}

\begin{figure}[ht!]
 \includegraphics[width=0.48\textwidth]{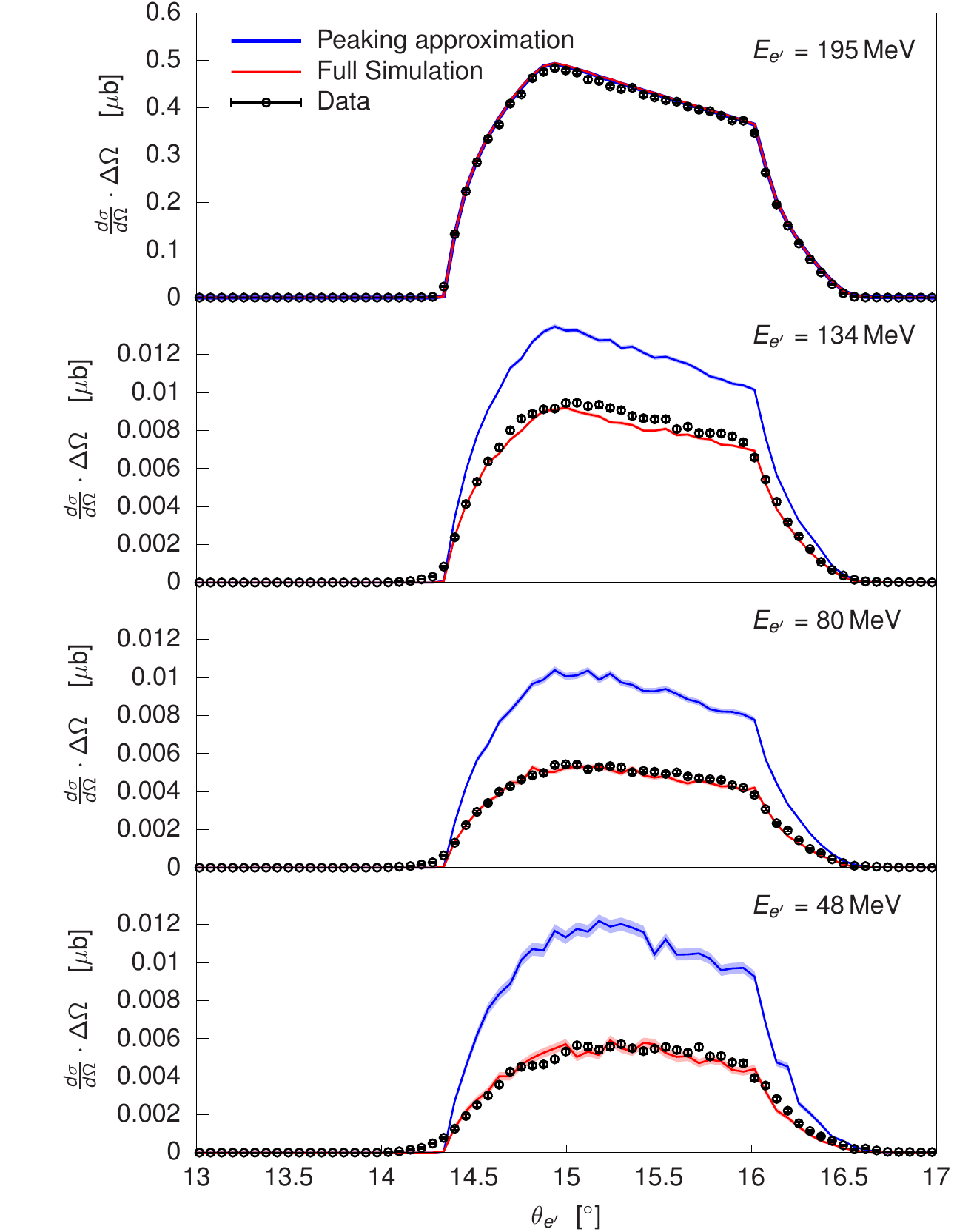}
 \caption{ The cross-sections for electron-carbon scattering as a function of scattering angle $\theta_{e'}$ at different values of energy of the detected electron $E_{e'}$. The measurements (black points) are presented together with the simulation based on the peaking approximation (blue line) and the simulation using the exact calculator of the Bethe-Heitler processes (red line).  
 \label{fig_ResultsAngles}}
\end{figure}

To simulate the acceptance averaged cross-sections we embedded the developed event generator into the standard simulation package of the A1 Collaboration, which includes a detailed description of the experimental apparatus~\cite{Blomqvist}. For each kinematic configuration the simulation was run for $2\cdot10^8$ events. The obtained distributions for $\Delta E_{e'}$ and $\theta_{e'}$ at various $E_{e'}$ are presented in  Fig.~\ref{fig_ResultsEnergy} and Fig.~\ref{fig_ResultsAngles}.

The relative statistical uncertainty of the collected data is $1.5\,\mathrm{\%}$ for every energy bin. This uncertainty needs to be considered together with the corresponding systematic uncertainty. Although the ISR experiment provided a remarkable control over the systematic uncertainties~\cite{Mihovilovic2019}, a few ambiguities remain and limit the precision of the comparison. These include: the experimental uncertainty related to the determination of the luminosity and detector efficiencies ($0.3\,\mathrm{\%}$); the contribution of higher-order corrections, which are not included in the simulation $(0.4\,\mathrm{\%})$; the uncertainty of the form-factor model ($1\,\%$); the uncertainty due to the omission of carbon pole diagrams $(0.03\,\mathrm{\%})$; use of the formalism of Mo and Tsai~\cite{MoTsai1969} to describe external corrections to the cross-section ($0.24\,\%$).


\section{Results and conclusions}
The new simulation described in Sec.~\ref{sec:fullsim} exhibits a very good overall agreement with the data. Fig.~\ref{fig_ResultsEnergy} shows that for $\Delta E_{e'} \leq 20\,\mathrm{MeV}$ the relative difference between the data and simulation is $1\,\mathrm{\%}$. For $\Delta E_{e'}$ between  $20\,\mathrm{MeV}$ and $150\,\mathrm{MeV}$ the agreement between the simulation and the data is between $5\,\mathrm{\%}$ and $10\,\mathrm{\%}$. 

The measurement of the cross-section $d\sigma/d\Omega_{e'}dE_{e'}$ was performed at smallest achievable beam energy $E_{e}$ and smallest reachable scattering angle $\theta_{e'}$ to minimize the contributions from the excited states of carbon as well as from quasi-elastic scattering processes that would appear as distortions in the measured distributions. While the excited states were indeed suppressed, the contributions of quasi-elastic (QE) events can be observed in the data in the region around  $E_{e'} = 172\,\mathrm{MeV}$. Since the significant discrepancy between the simulation and the data appears right at the QE peak and then decreases when moving away from it, we believe that the majority of the observed inconsistency arises from the contributions of QE peak and its corresponding radiative tail. Unfortunately this effect can not be efficiently added the simulation. 

Apart from that, the inconsistency between the data and simulation could arise from the incomplete description of higher-order corrections, especially the virtual vertex corrections and two real-photon emission diagrams. The virtual corrections to the Bethe-Heitler diagrams (self-energy corrections and various vertex corrections) require integration of the loop diagrams and are computationally very intensive, but could, in-principle, be considered by following the prescription of Ref.~\cite{Vanderhaeghen2000}. On the other hand, an inclusion of a better description of two-photon emission diagrams would represent a significant extension of the present model that could not be easily accomplished. 

In spite of the presented limitations, Fig.~\ref{fig_ResultsEnergy} and Fig.~\ref{fig_ResultsAngles} demonstrate that the new generator of events is superior to the traditionally used peaking approximation discussed in Sec.~\ref{sec:PA}. At $E_{e} = 195\,\mathrm{MeV}$ the approximation works reliably only for $\Delta E_{e'}\leq 1\,\mathrm{MeV}$. For other $\Delta E_{e'}$ the approximation overestimates the size of the radiative tail. At $\Delta E_{e'} \approx 10\,\mathrm{MeV}$ the discrepancy amounts to $+10\,\mathrm{\%}$, which then quickly increases to above $100\,\mathrm{\%}$ at $\Delta E_{e'} = 50\,\mathrm{MeV}$.

To conclude, we presented a new Monte-Carlo generator of events for mimicking QED radiative corrections in electron-carbon scattering process that is based on the exact calculation of first order Bethe-Heitler diagrams. Using a dedicated set of experimental data from MAMI together with the analysis package of the A1 Collaboration, we were able to test the generator and demonstrated that it can be successfully applied in the analyses in which a reliable subtraction of Bethe-Heitler background caused by the elastic peak is required. With the present generator's limitations the intensity of the studied inelastic process (e.g. low energy quasi-elastic scattering) should be at least two-times larger than the intensity of the radiative tail underneath.

\bmhead{Acknowledgments}

The authors would like to thank the MAMI accelerator group for the excellent beam quality which made this experiment possible. This work is supported by the Federal State of Rhineland-Palatinate, by the Deutsche Forschungsgemeinschaft with the Collaborative Research Center 1044, by the Slovenian Research Agency under Grants P1-0102 and J1-4383 and by Croatian Science Foundation under the project IP-2018-01-8570.

\bibliography{MihovilovicBibliography}

\end{document}